\documentclass[a4paper]{article}

\usepackage{INTERSPEECH2022}
\usepackage{subcaption}
\usepackage{makecell}
\usepackage{multirow}

\title{DeID-VC: Speaker De-identification via Zero-shot Pseudo Voice Conversion}
\name{Ruibin Yuan, Yuxuan Wu, Jacob Li, Jaxter Kim}
\address{
  Carnegie Mellon University, United States} 
\email{\{ruibiny,yuxuanw2,jacobli,jaxterk\}@andrew.cmu.edu}

\begin{document}

\maketitle
\begin{abstract} 
The widespread adoption of speech-based online services raises security and privacy concerns regarding the data that they use and share. If the data were compromised, attackers could exploit user speech to bypass speaker verification systems or even impersonate users. To mitigate this, we propose DeID-VC\footnote{Code and listening demo: https://github.com/a43992899/DeID-VC}, a speaker de-identification system that converts a real speaker to pseudo speakers, thus removing or obfuscating the speaker-dependent attributes from a spoken voice. The key components of DeID-VC include a Variational Autoencoder (VAE) based Pseudo Speaker Generator (PSG) and a voice conversion Autoencoder (AE) under zero-shot settings. With the help of PSG, DeID-VC can assign unique pseudo speakers at speaker level or even at utterance level. Also, two novel learning objectives are added to bridge the gap between training and inference of zero-shot voice conversion. We present our experimental results with word error rate (WER) and equal error rate (EER), along with three subjective metrics to evaluate the generated output of DeID-VC. The result shows that our method substantially improved intelligibility (WER 10\% lower) and de-identification effectiveness (EER 5\% higher) compared to our baseline.

\end{abstract}
\noindent\textbf{Index Terms}: privacy, anonymization, de-identification, voice conversion, pseudo speaker, speaker generation, Autoencoder, VAE, zero-shot.

\section{Introduction}
Many web services rely on automatic speaker verification (ASV) or automatic speech recognition (ASR) systems to verify user identities or transcribe user speech. These applications often require data sharing to cloud servers, which poses a major threat to user privacy and account security. An attacker with access to a user's voice data can utilize a text-to-speech system and a speaker embedding extractor to generate adversarial utterances that sound like they were spoken by the user. These phony utterances could be used to bypass ASV systems or to send scam voice messages to the user's friends and relatives \cite{kinnunen2012vulnerability}.

Speaker de-identification \cite{tomashenko2020introducing} serves as a way to prevent these illicit uses of speech data. This task involves removing unique speaker-dependent properties and preventing the speaker from being identified. Several approaches have been proposed for this task, such as obfuscation, encryption, and anonymization. Obfuscation \cite{8918913} modifies speech data to make it unrecoverable, whereas encryption approaches \cite{brasser18_interspeech} have limitations due to their computational complexity. The anonymization methods \cite{7472729, qian2017voicemask}, on the other hand, disguise the speaker's identity to the fullest extent while preserving other information and attributes. Our work is in the scope of anonymization. By adapting neural voice conversion (VC) methods to this task, we modify a given speech from a source speaker to match the vocal qualities of a target speaker.

To represent speaker-dependent features, previous neural VC-based de-identification systems typically use speaker embedding extracted by speaker recognition systems. They either rely on pre-defined real target speakers \cite{srivastava2020evaluating} or pseudo speaker embeddings that are generated by averaging real target embeddings \cite{tomashenko2022voiceprivacy, han2020voice, fang2019speaker, srivastava2020design}. The former can lead to a lack of speaker diversity, and thus may not be suitable for scenarios that require speaker uniqueness while concealing speaker identities. When performing online perturbation on large speaker databases, the latter is time-consuming because it involves distance measuring. It may also produce unrealistic speaker embeddings, which can lead to conversion failure.

In this paper, we propose a speaker de-identification model, DeID-VC, which is a zero-shot VC model based on AutoVC \cite{qian2019autovc}, with two novel learning objectives to bridge the gap between training and inference of zero-shot VC. The objectives work by introducing different source-target speaker pairs into training to simulate inference setting.
A novel VAE-based pseudo speaker generator (PSG) is also proposed, which is trained with a carefully designed learning objective. This objective enhances the convergence of the VAE and allows better modeling of speaker embeddings space. By sampling from noise, the PSG allows DeID-VC to produce an arbitrarily large number of realistic pseudo target embeddings in a short time. We evaluate DeID-VC using 5 different metrics: Word Error Rate (WER) of the model evaluated by a pretrained ASR model, Equal Error Rate (EER) of speaker verification, subjective naturalness, subjective intelligibility, and subjective verifiability. 

\section{Method}
\label{Method}
\subsection{Baseline Method}
\begin{figure}[th]
        \begin{subfigure}[c]{0.49\columnwidth}
            \centering
            \includegraphics[width=\columnwidth]{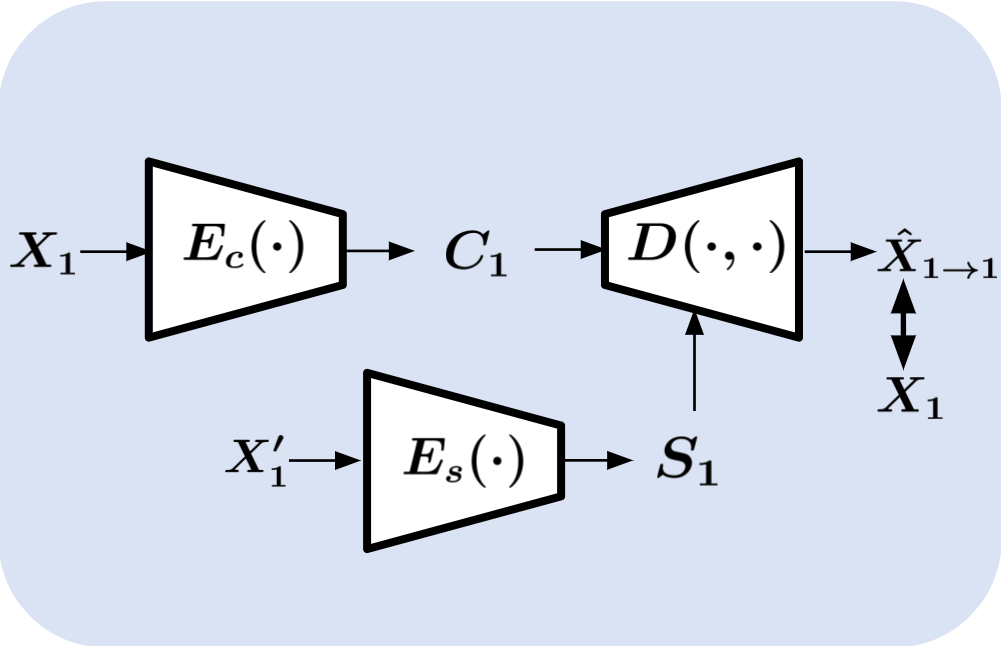}
            \captionof{figure}{Training Pipeline}
        \end{subfigure}\hfill
        \begin{subfigure}[c]{0.49\columnwidth}
            \centering
            \includegraphics[width=\columnwidth]{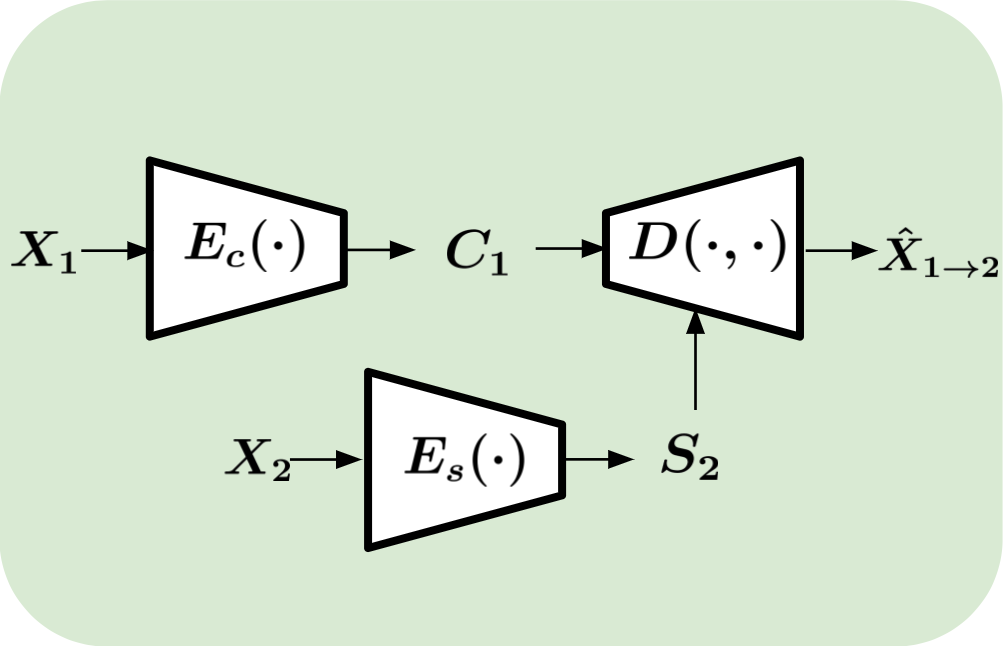}
            \captionof{figure}{Inference Pipeline}
        \end{subfigure}
        \caption{AutoVC as Baseline. A narrow bottleneck disentangles speaker style $S$ from speech content $C$. It contains a content encoder $E_c(\cdot)$, a speaker encoder $E_s(\cdot)$, and a decoder $D(\cdot,\cdot)$. (a) It is trained by minimizing self-reconstruction losses on the same source-target speaker pair. (b) The source-target pair is different during inference.}
        \label{fig:baseline figure}
    \end{figure}

    The baseline model we used is \textbf{AutoVC} \cite{qian2019autovc}, an autoencoder designed with a narrow information bottleneck that disentangles speaker style from speech content and converts source speech to the voice of target speaker. Fig.~\ref{fig:baseline figure} shows the training and inference pipeline for it. It contains three parts:
    
    (1) A content encoder $E_c(\cdot)$ that encodes $X_1$, the $80$-dimensional source mel-spectrogram sequence, into $C_1$, the $64$-dimensional content embedding sequence. The output is downsampled by 32 along the time axis. This narrow bottleneck is designed to squeeze out the speaker information from $C_1$. 
    
    (2) A D-Vector \cite{variani2014deep} based speaker encoder $E_s(\cdot)$ that extracts $S_1$ or $S_2$, the $256$-dimensional target speaker embeddings vector, from $X_1^{\prime}$ or $X_2$, the target mel-spectrogram sequence. The target speaker is the same as the source (subscript 1) during training, but different (subscript 2) during inference. Note that it is pre-trained on LibriSpeech \cite{panayotov2015librispeech} and VoxCeleb1 \cite{nagrani2020voxceleb} with GE2E loss \cite{speakerencoder} and fixed throughout training.
    
    (3) A decoder $D(\cdot,\cdot)$ that combines content embeddings and target speaker embeddings to generate converted mel-spectrograms. This also contains a post-net for output refinement, which predicts the residual of the mel-spectrograms. The output mel-spectrograms before and after refinement are denoted by $\hat{X}_{a\rightarrow b}$ and $\tilde{X}_{a\rightarrow b}$ respectively, where $a$ and $b$ are the source and target speaker identity respectively.

    The training objective $L$ is to minimize the reconstruction losses in terms of raw output, refined output, and content embedding, denoted by $L_{\text{recon}}$, $L_{\text{recon0}}$, $L_{\text{content}}$ respectively.
    \[\min_{E_c(\cdot),D(\cdot,\cdot)} L = L_{\text{recon}} + \mu L_{\text{recon0}} + \lambda L_{\text{content}}\]
    where
    \[L_{\text{recon}} = \mathbb{E}\left[\left\|\hat{X}_{1\rightarrow 1} - X_1\right\|^2_2\right] \]
    \[L_{\text{recon0}} = \mathbb{E}\left[\left\|\tilde{X}_{1\rightarrow 1} - X_1\right\|^2_2\right]\]
    \[L_{\text{content}} = \mathbb{E}\left[\left\|E_c(\hat{X}_{1\rightarrow 1}) - C_1\right\|_1\right]\]
    $\mathbb{E}[\cdot]$ denotes the expectation, and $\mu$ and $\lambda$ are the weights of corresponding objectives.

\subsection{Proposed Method}
    Our proposed method is called \textbf{DeID-VC}, an Autoencoder with a similar bottleneck architecture as AutoVC, but with new learning objectives and a well trained Pseudo Speaker Generator.
    
\subsubsection{Objectives for bridging the gap}
    In theory, an ideally sized information bottleneck can disentangle speech content $C$ from speaker style $S$. Therefore, once the bottleneck is properly set, the network only needs to focus on learning the self-reconstruction, that is, performing the voice conversion process where the source and target speakers are the same. However, the fixed length bottleneck is not always perfect, as the ideal dimension of the bottleneck can vary between different datasets, speakers, or utterances. Due to the leftover speaker identity information in $C$, training with self reconstruction objectives alone can sometimes lead to overfitting, because a network with large enough capacity can leverage this information to achieve good reconstruction results without fully utilizing $S$. 
    
    In preliminary experiments, a mismatch of performance during training and inference was observed, where noticeable intelligibility degradation can be found in AutoVC during inference, which is potentially the result of overfitting. Specifically, we call this performance mismatch a gap between training and inference.
    
\begin{figure}[th]
        \centering
        \includegraphics[width=\columnwidth]{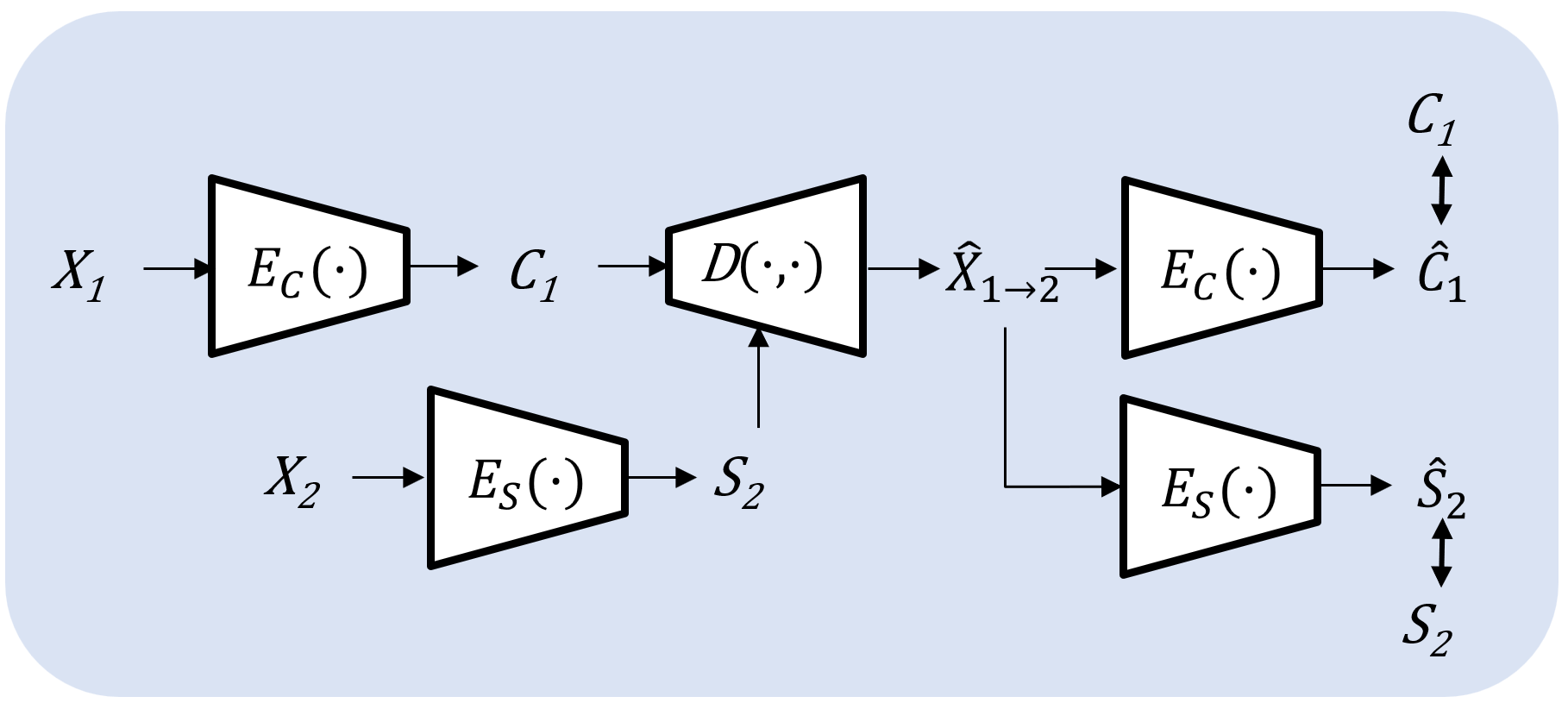}
        \caption{DeID-VC Training Pipeline. We incorporate different source-target speaker pairs into training to simulate inference setting. The model learns to preserve content and speaker consistency with these new speaker pairs.}
        \label{fig:proposed figure}
    \end{figure}

    To bridge this gap, we propose a new training pipeline shown in Fig.~\ref{fig:proposed figure}. We add two new learning objectives: a content consistency loss $L_{\text{content\_consist}}$ and a speaker consistency loss $L_{\text{speaker\_consist}}$. The key idea of these two losses is to simulate the inference setting, and better disentangle $C$ from $S$ by introducing different source and target speakers during training. The speaker consistency term aims to force the network to utilize $S$ by minimizing the $L1$ distance between the converted speaker embedding $\hat{S}_2$ and the target speaker embedding $S_2$. Similarly, the content consistency term works by forcing the network to utilize $C$, in which we minimize the $L1$ distance between the converted content embedding $\hat{C}_1$ and the source content embedding $C_1$.
    
    During training, we pass the source content embedding $C_1$ and the target speaker embedding $S_2$ through the decoder to get a converted mel-spectrogram $\hat{X}_{1\rightarrow 2}$. We then pass $\hat{X}_{1\rightarrow 2}$ through $E_{c}(\cdot)$ to get a prediction of $\hat{C}_1$, and also through $E_{s}(\cdot)$ to get a prediction of $\hat{S}_2$. 
    
    The overall training objective becomes the weighted summation of the training objectives in AutoVC and the two additional consistency losses, computed by:
    \begin{align*}
        \min_{E_{c}(\cdot), D(\cdot,\cdot)} L & = L_{\text{recon}} + \mu L_{\text{recon0}} + \lambda L_{\text{content}} \\ & \quad + \alpha L_{\text{content\_consist}} + \beta L_{\text{speaker\_consist}}
    \end{align*}
    where
    \[L_{\text{content\_consist}} = \mathbb{E}\left[\left\lvert E_{c} (\hat{X}_{1\rightarrow 2}) - C_{1}\right\rvert_{1}\right]\]
    \[L_{\text{speaker\_consist}} = \mathbb{E}\left[\left\lvert E_{s} (\hat{X}_{1\rightarrow 2}) - S_{2}\right\rvert_{1}\right]\]
    $\alpha$ and $\beta$ are the weights of corresponding objectives.

\subsubsection{Pseudo speaker generator}
\label{PSG}
    Most of the existing voice conversion systems use a limited pool of pre-defined target speaker models based on real speakers. Even though our system can perform zero-shot voice conversion, using real speakers for target speaker models limits its capability. As a result, any limit on the number of real speakers an implementation can access would similarly limit our target speaker diversity. In practice, there may also sometimes be a need to preserve some level of speaker uniqueness while obfuscating the real source speaker voices.
    
\begin{figure}[th]
        \begin{subfigure}[c]{0.49\columnwidth}
            \centering
            \includegraphics[width=\columnwidth]{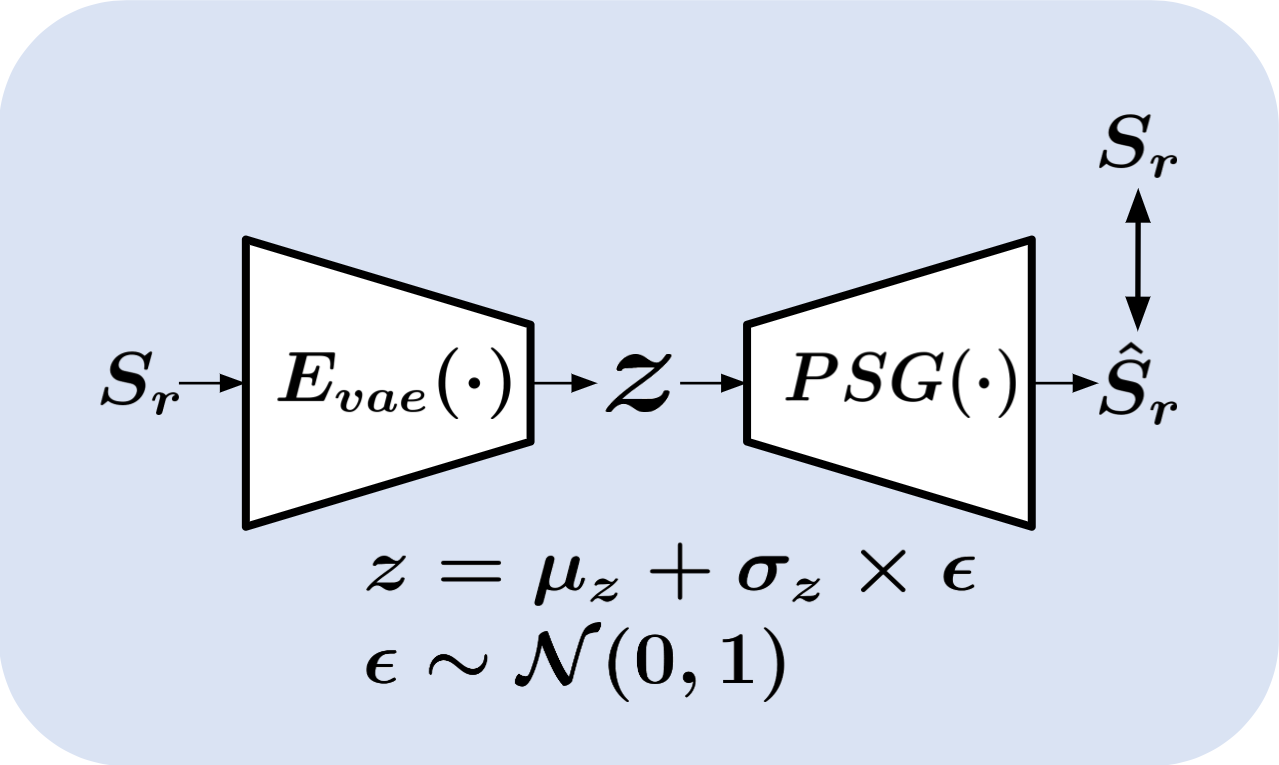}
            \captionof{figure}{VAE Pretraining}
            \label{fig:PSG train figure}
        \end{subfigure}\hfill
        \begin{subfigure}[c]{0.49\columnwidth}
            \centering
            \includegraphics[width=\columnwidth]{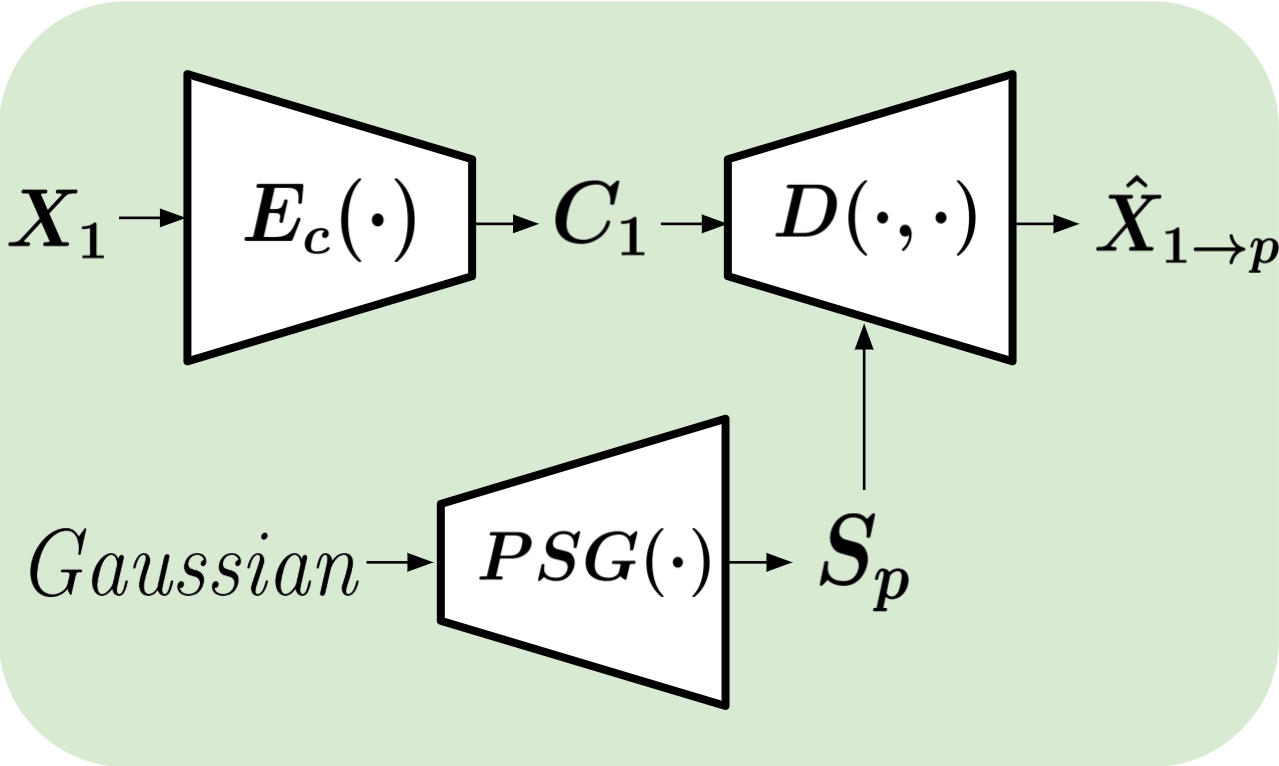}
            \captionof{figure}{DeID-VC Inference with PSG}
            \label{fig:PSG infer figure}
        \end{subfigure}
        \caption{Pseudo Speaker Generator. (a) It is pretrained in a VAE setting with a new learning objective. (b) Simply replacing $E_s(\cdot)$ with $PSG(\cdot)$ during inference can enable fast, high quality, and unlimited pseudo speaker generation.}
        \label{fig:PSG figure}
    \end{figure}

    To overcome this problem, we design a VAE \cite{kingma2013auto} based \textbf{Pseudo Speaker Generator} (PSG) to generate an unlimited amount of pseudo speaker embeddings $S_p$ from Gaussian noise. The PSG module is pretrained separately as the decoder of a VAE, which learns to reconstruct real speaker embeddings by mapping the real data distribution to a Gaussian distribution. The pretraining of the VAE and the inference pipeline using PSG is shown in Fig.~\ref{fig:PSG figure}. 
    
    During generation, simply replacing the speaker encoder with our PSG module can enable a flexible level and amount of target speaker assignment. For example, the source speaker's voice can be de-identified in utterance level before uploading to the cloud, by converting each utterance to pseudo target voices $S_p$. In another case, user-level speaker assignment is also possible, that is, assigning a unique target voice to each user ID.
    
    For the training part, previous works \cite{zhang2019vae, wang2019vae} tend to use deep and wide components to train VAE on speaker embeddings. However, we found that this is unnecessary, as shallow and narrow components can also do the trick. A simpler structure leads to less structural risk and enables the possibility to deploy the PSG module to edge devices. To achieve this, we carefully design a learning objective for efficient convergence and better modeling of the speaker embedding space. It has three components: an L1 reconstruction term, a cosine distance term, and a KL regularization term.
    
    In the field of computer vision, studies \cite{zhao2016loss, wang2009mean} have discovered that using $L2$ loss on image reconstruction tasks can lead to unsatisfactory blurry results. Part of the reason is that $L2$ loss is bad at handling small errors, as the magnitude of the gradient is not sufficient when the predictions and targets are close \cite{feng2018wing}. The task of speaker embedding reconstruction suffers from a similar problem, as the residual between the reconstructed embedding $\hat{S}_{r}$ and the source embedding $S_{r}$ is small. Therefore, to improve the convergence of our model, we replace the commonly used $L2$ loss with $L1$ loss, which is computed by:
    \[L1 = \sum \left| S_{r} - \hat{S}_{r} \right|\]
    
    However, both $L1$ and $L2$ loss do not ensure the network learns to model the speaker embedding space properly. For systems like D-Vector that use cosine similarity to evaluate the speaker identities, the angle between $\hat{S}_{r}$ and $S_{r}$ should be as small as possible. To preserve the speaker's identity after reconstruction, we add a cosine distance term, computed by:
    \[L_{\text{dist}} = \sum (1 - \dfrac{S_{r} \cdot \hat{S}_{r}}{\Vert S_{r} \Vert _2 \cdot \Vert \hat{S}_{r} \Vert _2})\]
    
    Finally, just as in a conventional VAE, a KL-divergence term is preserved and computed by:
    \[L_{\text{kl}} = KL\Big(\mathcal{N}(\mu_z, \sigma_z^2) || \mathcal{N}(0, 1)\Big) = \frac{1}{2} \sum (\mu_z^{2} + \sigma_z^{2} - \log{\sigma_z^{2} - 1)}\] 
    where $\mu_z$ and $\sigma_z^2$ is the mean and covariance of the encoded latent distribution, from which the latent vector $z$ is sampled: 
    \[z = \mu_z + \sigma_z \times \epsilon, \quad
    \epsilon \sim \mathcal{N}(0, 1)\]

    The proposed objective for VAE pretraining is then the weighted sum of the above three components. $\lambda_\text{dist}$ is the weight of $L_\text{dist}$.
    \begin{align*}
        \min_{E_{vae}(\cdot), PSG(\cdot)} L & = L1 + \lambda_\text{dist} L_\text{dist} + L_{\text{kl}}
    \end{align*}
    
\subsubsection{Spectrogram Inverter}
    In AutoVC, WaveNet \cite{oord2016wavenet} is used as the vocoder to reconstruct audio samples from mel-spectrograms. However, it takes around 2-3 minutes to generate 5 seconds of audio on a GPU. 
    In order to speed up the conversion process, we replace WaveNet with \textbf{HiFi-GAN} \cite{hifigan} pre-trained on the VCTK corpus \cite{yamagishi2019cstr} as the vocoder. It ensures the fidelity and efficiency (50,000x faster than WaveNet) in the voice conversion task. In our results, we use HiFi-GAN for the vocoder in both the baseline and our method.

\section{Experiment}
\label{Experiment}

\subsection{Dataset}
    We train and report our baseline and proposed DeID-VC on the Wall Street Journal (WSJ) Corpus. It is the combination of WSJ0 \cite{wsj0} and WSJ1 \cite{wsj1}. We allocated 60\% of the speakers for training and the remaining 40\% for testing. 
    For the VAE based PSG pre-training part, we use the combination of VoxCeleb 1\&2 for training, VCTK and WSJ for validating.

\subsection{Training Setting}
    We split the DeID-VC training into two stages. At stage one, we follow the settings from AutoVC and train the baseline model for 500k steps until it converges, where we set $\mu$=1, $\lambda$=1. At stage two, we add in the two new objectives and fine-tune the stage one model for 10k steps, where we set $\mu$=1, $\lambda$=10, $\alpha$=10, $\beta$=0.1. For VAE based PSG pretraining, we set $\lambda_\text{dist}=200$. We use Adam optimizer and batch size of 128 for all training procedures. Note that finetuning the PSG module on WSJ training data can boost the conversion performance, the result shown in Table~\ref{tab:result} is after finetuning.
\begin{center}
\begin{table*}[]
  \centering
  \caption{Our Methods Compared to Baseline.}
  \label{tab:result}
    \begin{tabular}{c|cc|cccc|cccc}
        \hline
                                    & \multicolumn{2}{c|}{Baseline} & \multicolumn{4}{c|}{$+L_{\text{content\_consist}}$}               & \multicolumn{4}{c}{$+L_{\text{content\_consist}}+L_{\text{speaker\_consist}}$} \\
                                    & SxU                & UxU      & SxU            & UxU            & SxP            & UxP            & SxU         & UxU                  & SxP                 & UxP                 \\ \hline
        WER(\%)$\downarrow^+$             & 42.5              & 48.4    & \textbf{32.8} & \textbf{35.2} & \textbf{30.5} & \textbf{35.7} & 33.3       & 39.9                & 35.9               & 38.4               \\
        EER(\%)$\uparrow^+$               & \textbf{41.8}     & 36.1    & 40.7          & 36.7          & 41.0          & 37.6          & 41.7       & \textbf{38.4}       & \textbf{46.7}      & \textbf{38.3}      \\ \hline
        Intelligibility$\uparrow^+$   & 3.83               & 2.97     & \textbf{4.22}  & \textbf{3.72}  & \textbf{4.09}  & \textbf{3.84}  & 3.94        & 3.50                 & 2.81                & 2.78                \\
        Verifiability$\downarrow^+$   & 1.88               & 1.63     & \textbf{1.75}  & 1.66           & 1.78           & 1.66           & 1.81        & \textbf{1.59}        & \textbf{1.50}       & \textbf{1.47}       \\
        Naturalness$\uparrow^+$      & 3.06               & 3.03     & \textbf{4.06}  & \textbf{3.44}  & \textbf{3.63}  & \textbf{3.38}  & 3.56        & 2.78                 & 2.72                & 2.34                \\ \hline
    \end{tabular}
\end{table*}
\end{center}

\subsection{Evaluation Metrics}
    Three types of metrics are introduced for evaluating the converted speech: 
    
    (1) \textbf{ASR Word Error Rate (WER)}: To assess the intelligibility, we employed a Transformer based ASR model from ESPNet \cite{watanabe2018espnet} pre-trained on WSJ corpus, which reaches 6.6\% WER on WSJ test\_eavl92. 
    
    (2) \textbf{ASV Equal Error Rate (EER)}: To assess the ability of preventing attacks from speaker identification systems, we use ECAPA-TDNN \cite{desplanques2020ecapa} from Speechbrain \cite{speechbrain}, a state-of-the-art ASV model, which is pre-trained on VoxCeleb1+2 and reaches 0.69\% EER on VoxCeleb1-test and 0.16\% on WSJ corpus, as our adversary. The objective of our de-identification system is to confuse this model as much as possible. A higher EER indicates a better de-identification performance. 
    
    (3) \textbf{Mean Opinion Score}: Subjective evaluation is also adopted as the complement of the above metrics. Scores are given in three dimensions, including intelligibility, verifiability, and naturalness. All three scores range from 1 to 5. Scoring a speech 1 represents that the speech is totally unrecognizable as words, totally unverifiable comparing to the original speaker, or too unnatural in articulation. Inversely, scoring a speech 5 represents that the speech is clear and fully recognizable, obviously from the same speaker as the original, or is as natural as spoken from a real human. Surveys are conducted among 30 people with proficiency in English, and the mean scores are calculated afterwards.

\subsection{Results}
In order to simulate the de-identification process under real settings, we organize our experiments into four different testing scenarios: \textbf{Seen Source to Unseen Targets (SxU)}, \textbf{Unseen Source to Unseen Targets (UxU)}, \textbf{Seen Source to Pseudo Targets (SxP)}, \textbf{Unseen Source to Pseudo Targets (UxP)}. A seen source is an average speaker embedding extracted from 10 randomly selected 2-second utterances of a same speaker in the train set, while an unseen source is an average speaker embedding of a speaker in the test set. Similarly, we define an unseen target speaker here as the average speaker embedding extracted from 10 randomly selected 2-second utterances of a speaker in the test set. A pseudo target is a speaker embedding randomly generated by PSG. This experiment setting is intended to fully evaluate the performance of our model when dealing with unknown target speakers, as well as pseudo speakers.

We carry out our experiments with all three models, including the baseline model, the model finetuned with $L_{\text{content\_consist}}$, and the model finetuned with $L_{\text{content\_consist}}+L_{\text{speaker\_consist}}$. Note that during evaluation, we use utterance level target assignment, which means that different utterances use different target embeddings. The source and target pairs within a testing scenario are kept the same across different models.

As shown in Table~\ref{tab:result}, our model gets the lowest WER in all four scenarios when trained with content consistency loss. Training with both loss functions gives a slightly less satisfying result, but still significantly outperforms the baseline model. This indicates an improved intelligibility of our models, and the subjective intelligibility scores also support this conclusion.

The model trained with both losses got the highest EER and lowest verifiability score in three of the four scenarios, and the model trained with only the content consistency loss also outperforms or reaches comparable results to the baseline. This means our models, especially the one using the speaker consistency loss, are able to further improve the anonymizing ability of the baseline, when slightly sacrificing the intelligibility and naturalness. Note that when applying pseudo targets, the intelligibility and naturalness metrics scores are similar to using real targets, which indicates that the PSG module is able to generate realistic pseudo targets. 

Generally the results on seen sources have better performance in both intelligibility and verifiability. The presence of unseen speakers slightly deteriorates the results, but still outperforms the baseline. This fact shows that our models are still fairly effective even in zero-shot scenarios.

\subsection{PSG Analysis}
\label{PSG Analysis}
We use a single hidden layer MLP architecture for both the VAE encoder and PSG. The hidden size is 384 and the latent dimension is 64. The learning rate is set to 1e-3 and the network is trained for 60 epochs. To measure the convergence of the pretrained model and whether it learns to capture the geometric property of the embeddings, we compare the reconstructed speaker embeddings with the source embeddings on the training and validation set, in terms of Mean Square Error (MSE) and Cosine Similarity (CosSim). As indicated in Table~\ref{tab:vae eval}, $L2$ loss does not guarantee the CosSim after reconstruction and the MSE is higher than the model trained with $L1$ loss. Adding $L_{\text{dist}}$ to $L2$ can significantly improve the CosSim, but harms the MSE. However, combining $L1$ with $L_{\text{dist}}$ can not only reduce the MSE but also improve the CosSim, thus providing the best result. This supports our arguments in \ref{PSG}, and shows the effectiveness of the proposed objective for learning a VAE-based PSG.

\begin{table}[]
\caption{Reconstruction Measurement of the VAE.}
\label{tab:vae eval}

\begin{tabular}{ccccc}
\hline
                                                     &                                      & $L2$     & $L2+L_{\text{dist}}$ & $L1+L_{\text{dist}}$        \\ \hline
\multicolumn{1}{c|}{\multirow{3}{*}{MSE$\downarrow^+$}}            & \multicolumn{1}{c|}{Vox(train)} & 0.4044 & 0.4072   & \textbf{0.0831} \\
\multicolumn{1}{c|}{}                                & \multicolumn{1}{c|}{VCTK(val)}       & 0.6284 & 0.9691   & \textbf{0.5267} \\
\multicolumn{1}{c|}{}                                & \multicolumn{1}{c|}{WSJ(val)}        & 0.6278 & 0.8421   & \textbf{0.5122} \\ \hline
\multicolumn{1}{c|}{\multirow{3}{*}{CosSim$\uparrow^+$}} & \multicolumn{1}{c|}{Vox(train)} & 0.1892 & 0.8698   & \textbf{0.9164} \\
\multicolumn{1}{c|}{}                                & \multicolumn{1}{c|}{VCTK(val)}       & 0.0905 & 0.4446   & \textbf{0.4972} \\
\multicolumn{1}{c|}{}                                & \multicolumn{1}{c|}{WSJ(val)}        & 0.0557 & 0.4448   & \textbf{0.5297} \\ \hline
\end{tabular}
\end{table}

\section{Conclusion \& Future Work}
\label{Conclusion}
In this paper, we proposed DeID-VC, a speaker de-identification model that improves the training procedure of AutoVC with two additional loss terms, along with a pseudo speaker generator that generates high quality pseudo targets. In preliminary experiments, we discovered that $L2$ loss may not be a reliable measure of speech reconstruction quality, therefore the training can be unstable and fall into a bad local minimum with poor conversion intelligibility. In addition, training with more data can enhance intelligibility significantly. We may experiment with large-scale data in the future and look at better reconstruction objectives, such as perceptual loss.


\bibliographystyle{IEEEtran}

\bibliography{final}

\end{document}